\documentclass{article}

 \usepackage[ninepoint]{eusipco2011}          
\usepackage{amsmath,amssymb}
\usepackage{graphicx}
\newcommand\defin[1]{\textbf{Definition #1:}}
\title{OFDM Pilot Allocation for Sparse Channel Estimation}

\name {Pooria Pakrooh, Arash Amini, and Farrokh Marvasti}

\address{ Advanced Communication Research Institute (ACRI)\\
EE Department, Sharif University of Technology, Tehran, Iran\\
\{pakrooh,arashsil\}@ee.sharif.edu, and marvasti@sharif.edu}

\begin{document}

\maketitle

\begin{abstract}
In communication systems, efficient use of the spectrum is an indispensable concern. Recently the use of compressed sensing for the purpose of estimating Orthogonal Frequency Division Multiplexing (OFDM) sparse multipath channels has been proposed to decrease the transmitted overhead in form of the pilot subcarriers which are essential for channel estimation. In this paper, we investigate the problem of deterministic pilot allocation in OFDM systems. The method is based on minimizing the coherence of the submatrix of the unitary Discrete Fourier Transform (DFT) matrix associated with the pilot subcarriers. Unlike the usual case of equidistant pilot subcarriers, we show that non-uniform patterns based on cyclic difference sets are optimal. In cases where there are no difference sets, we perform a greedy method for finding a suboptimal solution. We also investigate the performance of the recovery methods such as Orthogonal Matching Pursuit (OMP) and Iterative Method with Adaptive Thresholding (IMAT) for estimation of the channel taps.\\

\end{abstract}


\section{Introduction}

In wireless communications, Orthogonal Frequency Division Multiplexing (OFDM) is a well-known solution for overcoming the problem of multipath fading channels \cite{ar1}, \cite{ar2}. However, this solution is effective only when the receiver is provided with tools to estimate the Channel Frequency Response (CFR). To this end, the transmitter should send some predefined data in a predefined order that the receiver is a priori aware of. These predefined data are usually called pilots.

There are two main approaches for inserting pilot data in OFDM signals. In block-type pilots, all the subcarriers in some OFDM blocks (the whole spectrum) are reserved as pilot tones. In comb-type pilot models, some predefined subcarriers in each block serve as pilots. Hence, CFR at these subcarriers can be estimated using methods such as Least Square (LS) or Minimum Mean Square Error (MMSE). Now for estimating the CFR at non-pilot subcarriers, interpolation methods ranging from simple linear or second order techniques \cite{4-bahayi} to time domain \cite{5-bahayi} and even more complex approaches are used. It is clear that by decreasing the frequency gap between the adjacent pilot subcarriers, the performance of the interpolation techniques improves. Therefore, the pilots are preferably put at equidistant subcarriers to provide uniformity.

Considering the inherent sparsity in the impulse response of the wireless channels which is due to the sparse structure of the scattering objects, it is possible to estimate the Channel Impulse Response (CIR) more accurately even from non-uniform pilot patterns. The common estimation techniques in this case are those introduced in the field of compressed sensing such as basis pursuit \cite{candes} and OMP \cite{greedisgood}. Unlike the interpolation case, equidistant pilot locations are not the best choices here. In \cite{taubock}, using the results of \cite{candes} for sparse signal recovery, it is mentioned that uniformly random pilot locations\footnote{That means all possible choices of pilot indices are equally likely} can provide the possibility of perfect channel reconstruction with overwhelming probability. Although this is an important theoretical result, it is not practical. In this paper, we suggest a deterministic structure for the pilot locations in sparsity-based channel estimation methods which minimizes the inter-atom interference in DFT submatrices. Simulation results confirm the efficiency of the proposed pilot allocation method when greedy methods are used for channel estimation. Also, we propose an iterative thresholding method for channel estimation which results in appropriate performance in time-variant frequency selective OFDM channels.

\section{Problem Statement}

In OFDM systems with comb-type pilot arrangement, ignoring the effects of Inter-Symbol Interference (ISI) and Inter-Carrier Interference (ICI)
, the received data at the $k^{th}$ subcarrier ($1\leq k\leq N$) of the $n^{th}$ OFDM frame can be formulated as:
\begin{equation}
 Y(n,k)=X(n,k)\cdot H(n,k)+W(n,k),
\end{equation}
where $X(n,k)$ is the transmitted OFDM symbol, $H(n,k)$ is the channel frequency response and $W(n,k)$ is the AWGN noise.
If $\mathcal{P}$ denotes the set of all pilot indices, at a given pilot subcarriers $k_p\in \mathcal{P}$ and using the least square method, the CFR can be estimated as:
\begin{equation}
\tilde{H}(n,k_p)=\frac{Y(n,k_p)}{X(n,k_p)}=H(n,k_p)+\frac{W(n,k_p)}{X(n,k_p)}.
\end{equation}

As explained earlier, conventional methods for estimation of the CFR at non-pilot subcarriers (given the noisy measurements at pilots) are interpolation-based techniques which require relatively high sampling rates (number of pilots) to produce acceptable Mean Squared Error (MSE).
Also, the optimum structure of the pilot locations for these techniques which minimizes the MSE of the estimated channel, is the uniform distribution (equidistance) of the pilots in the spectrum.

In the sparsity-based channel estimation methods, instead of finding the CFR, the goal is to estimate the inherently sparse CIR in each OFDM frame from limited number of noisy measurements of the CFR obtained at pilot locations. The estimated CIR is then, translated into the frequency domain by means of FFT which results in an estimation of the CFR that can be used for data equalization process. In these methods, we are dealing with the following system of equations:
\begin{equation}
\label{eq:3}
\tilde{\mathbf{H}}_p=\mathbf{F}_p\cdot \mathbf{h}+\mathbf{n}_p,
\end{equation}
where $\mathbf{F}_p$ is the DFT submatrix with $N_p=|\mathcal{P}|$ rows associated with the pilot locations, $\tilde{\mathbf{H}}_p$ is the vector of LS-estimated CFR at pilot locations, $\mathbf{h}$ is the sparse CIR vector, and $\mathbf{n}_p$ is the vector of noise values. 

Generally, there are two main categories of sparsity-based methods to solve the set of equations presented in (\ref{eq:3}). One approach is to minimize the $\ell_1$ norm of $\mathbf{h}$ subject to (\ref{eq:3}), either directly or iteratively (such as SPGL \cite{SPGL1}). Although the performance of such methods are considered among the bests, they are extremely slow for real-time implementation. The other approach which is considered in this paper, is to use fast greedy methods such as OMP which iteratively detect and estimate the location and value of the channel taps. These methods are usually faster than $\ell_1$ minimization techniques by orders of magnitude while they may fall short of performance. Our simulation results confirm that their performance is acceptable for the purpose of OFDM channel estimation.

The main advantages of sparsity-based approaches can be categorized into two parts:

\textbf{(1) Decreasing MSE:} Generally, the purpose of using compressed sensing methods in solving a linear set of equations with the sparsity constraint is to achieve the Cramer Rao lower bound on MSE \cite{2-Nd}. In extreme cases, the best estimator which achieves this lower bound is the structural LS estimator (usually called oracle estimator) which exactly knows the location of nonzero taps (support) and estimates their corresponding values using LS \cite{2-Nd}. However, in general, there is no information about the location of the nonzero coefficients of $\mathbf{h}$ at the receiver and the structural LS estimator is not realizable. Simulation results indicate that we can get close to this bound by using proper sparsity-based methods.

\textbf{(2) Reducing Overhead:} Although the pilot subcarriers occupy a fraction of the spectrum, they do not convey any data. By reducing the number of pilot subcarriers, we increase the utilization efficiency of the spectrum while we may degrade the performance of the channel estimation block. As mentioned in \cite{taubock}, by considering the sparsity of the CIR, it is possible to capture the necessary information in the frequency domain in fewer number of pilots. The results in \cite{candes} show that $\ell_1$ minimization technique almost perfectly reconstructs the sparse CIR from (\ref{eq:3}) when the number of pilots is proportional to the number of channel taps. Furthermore, the reconstruction performance is independent of the location and value of the taps; i.e., unlike the interpolation-based methods, the number of required pilot subcarriers does not depend on the delay spread and degree of frequency selectivity of the channel.


\section{Iterative Thresholding Method for Sparse Channel Estimation}\label{sec:IMAT}

In this section, we propose an Iterative Method with Adaptive Thresholding (IMAT)\cite{IMAT} for the purpose of estimating the sparse CIR. In other words, we aim to identify non-zero channel taps and estimate their corresponding values using IMAT.

\begin{figure}
\centering
\includegraphics[totalheight=1.7cm,width=8cm]{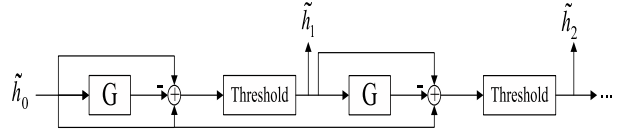}
\caption{Block diagram of the IMAT method}
\label{fig:iterative}
\end{figure}

In our general OFDM channel estimation problem presented in (\ref{eq:3}), our main goal is to estimate $\mathbf{h}$ from $\tilde{\mathbf{H}}_p$ given the fact that $\mathbf{h}$ has a few non-zero coefficients. To obtain an initial estimate, we multiply the sides of (\ref{eq:3}) by Moore-Penrose pseudo-inverse of $\mathbf{F}_{p}$ to find the solution with minimum $\ell_2$-norm:
\begin{equation}
\label{eq:5}
\tilde{\mathbf{h}}_0={\mathbf{F}_{p}}^\dagger\tilde{\mathbf{H}}_p ={\mathbf{F}_{p}}^\dagger \mathbf{F}_{p} \mathbf{h}+{\mathbf{F}_{p}}^\dagger \mathbf{n}_p.
\end{equation}
Using the properties of the Moore-Penrose psuedo-inverse for underdetermind set of equations we have:
\begin{equation}
{\mathbf{F}_{p}}^\dagger={\mathbf{F}_{p}}^H\underbrace{{(\mathbf{F}_{p}{\mathbf{F}_{p}}^H)}^{-1}}_{\frac{1}{N}\mathbf{I}_{N_P\times N_p}}=\frac{1}{N}{\mathbf{F}_{p}}^H.
\end{equation}
Now we can rewrite (\ref{eq:5}) as:
\begin{equation} \label{eq:8}
\tilde{\mathbf{h}}_0 =\underbrace{\frac{1}{N}{\mathbf{F}_{p}}^H\mathbf{F}_{p}}_{\mathbf{G}_{N\times N}}h+\frac{1}{N}{\mathbf{F}_{p}}^H\mathbf{n}_p.
\end{equation}
The elements of the $N\times N$ non-negative matrix $\mathbf{G}$ which is usually referred to as the distorting matrix, are given by:
\begin{equation}
g_{i,j}=\left\{
\begin{array}{cc}
\frac{1}{N} \langle\mathbf{f}_i\;,\;\mathbf{f}_j\rangle & i\neq j\\
 \\
\frac{N_p}{N} &  i=j
\end{array}\right. ,
\end{equation}
where $\mathbf{f}_i$ represents the $i^{th}$ column of $\mathbf{F}_p$. 
If the columns of $\mathbf{F}_p$ are orthogonal and there is no additive noise, $\tilde{\mathbf{h}}_0$ will be a scaled version of $\mathbf{h}$; in general case where the columns are not orthogonal, $\tilde{\mathbf{h}}_0$ is a distorted estimate of $\mathbf{h}$. Now through a series of iterations and by employing the sparsity constraint, we try to improve this estimate. In each iteration, we perform one step of the iterative method studied in \cite{marvasti} followed by a thresholding operator:
\begin{eqnarray}\label{eq:marvasti}
\tilde{\tilde{\mathbf{h}}}_{k}&=&\lambda (\tilde{\mathbf{h}}_0-\mathbf{G}\cdot \tilde{\mathbf{h}}_{k-1})+\tilde{\mathbf{h}}_{k-1}, \\
\tilde{h}_{k}(i)&=&\left\{
\begin{array}{cc}
\tilde{\tilde{h}}_{k}(i)& \,  |\tilde{\tilde{h}}_{k}(i)|>\beta e^{-\alpha k}\\
\\
0& \text{otherwise}
\end{array}\right.\label{eq:9},
\end{eqnarray}
where $\lambda$ and $k$ are the relaxation parameter and the iteration number, respectively, $\mathbf{G}\cdot {\tilde{\mathbf{h}}}_{k}$ is the output of the distorting operator to the input ${\tilde{\mathbf{h}}}_{k}$ and $\tilde{\mathbf{h}}_0$ is defined in (\ref{eq:8}). The steps (\ref{eq:marvasti}) are known to compensate for the non-orthogonality of the columns of $\mathbf{F}_p$ while the thresholding operator takes the sparsity constraint into account. We employ adaptive thresholds in (\ref{eq:9}) which can be tuned through the parameters $\alpha$ and $\beta$, and decrease exponentially with respect to the iteration number. 
The optimality of the exponential function in our method can be derived in a similar manner as in \cite{arash}. The block diagram of the proposed channel estimation method is shown in Fig. \ref{fig:iterative}.

In the proposed method, the ideal but unrealistic case would be when there is no distortion; i.e., when $\mathbf{G}$ is a scaled version of the identity matrix. Although this never happens, it shows that when $\mathbf{G}$ is a good-enough approximation of the identity matrix, we can expect satisfying results. Thus, a given set of pilot locations is considered as good if the off-diagonal elements of the matrix $\mathbf{G}$ are relatively small compared to the diagonal elements. In the next section, we will investigate the problem of selecting the DFT submatrix $\mathbf{F}_p$ which results in a suitable distorting matrix $\mathbf{G}$.


\section{Pilot Allocation by Minimizing the coherence in Partial DFT Matrices}\label{sec:Pilot_coherence}

As mentioned before, the performance of a channel estimation block depends on both the reconstruction technique and the set of pilot locations. In this section, we will study the sub-optimum pilot locations when greedy sparsity-based methods such as OMP and the introduced IMAT are employed.


\subsection{Cyclic Difference Sets: Minimum Coherence}\label{sec:DS}
To begin with, consider the following underdetermined set of equations:
\begin {equation}
\mathbf{y}_{m\times 1}=\mathbf{\Phi}_{m\times n} \mathbf{x}_{n\times 1}
\end {equation}
where $\mathbf{y}$ is the observed vector, $\mathbf{x}$ is an $s$-sparse vector (contains at most $s$ non-zero elements) and $\mathbf{\Phi}$ is an $m\times n$  ($m\ll n$) measurement matrix which in our case is the partial DFT matrix formed by selecting $N_p$ rows. In this section, we seek to find a proper location for pilots in each OFDM block, using the following definition.\\

\defin{1} The coherence of a measurement matrix $\mathbf{\Phi} \in \mathbb{C}^{m\times n}$ is the maximum absolute cross-correlation between the normalized columns:
\begin{equation}
\label{eq:11}
\mu_{\mathbf{\Phi}}=\max_{\substack{1\leq l,k \leq n \\ l\neq m}}  
\frac{|\langle\mathbf{\varphi}_l \;,\; \mathbf{\varphi}_k\rangle|}{\|\varphi_l\|_2\cdot \|\varphi_k\|_2} .
\end{equation}

Although the so called Restricted Isometry Property (RIP) \cite{candes} is the best known tool for characterizing the performance of a given matrix $\mathbf{\Phi}$ in sampling the sparse vectors, there is currently no polynomial time algorithm to check this property \cite{coeldar}. The common alternative for measuring how well a matrix preserves the information of the sparse vectors ($\mathbf{x}$) in the produced samples ($\mathbf{y}$) is the coherence; the smaller the better. In addition, the performance of the greedy methods is more influenced by the coherence of the measurement matrix rather than its RIP order \cite{greedisgood}. One of the well-known results demonstrates that the sparsity-based methods such as $\ell_1$ minimization and greedy methods, are guaranteed to perfectly recover the $s$-sparse vectors when  $\mu_{\mathbf{\Phi}} < \frac{1}{2s}$ \cite{greedisgood}.


Returning to our main problem stated in (\ref{eq:3}), we aim to choose pilot indices in such a way that the coherence of the resulted measurement matrix, $\mathbf{F}_p$, becomes as small as possible. Considering the unit-norm property of the elements of $\mathbf{F}_p$, we have:
\begin{equation}
\label{eq:12}
\mu_{\mathbf{F}_P}=\max_{\substack{1\leq l,k\leq n \\ l\neq k}}   \frac{|\langle \mathbf{f}_l \;,\; \mathbf{f}_k\rangle|}{N_p}.
\end{equation}

If the pilot indices are $\mathcal{P}= \{ P_1,\dots,P_{N_P}\}$, $\mathbf{F}_p$ becomes:
\begin{equation}
\mathbf{F}_p=\left[\begin{array}{cccc}
1&e^{-j\frac{2\pi}{N} P_1}&\ldots&e^{-j\frac{2\pi}{N} P_1(N-1)}\\
1&e^{-j\frac{2\pi}{N} P_2}&\ldots&e^{-j\frac{2\pi}{N} P_2(N-1)}\\
\vdots&\vdots&\vdots&\vdots\\
1&e^{-j\frac{2\pi}{N} P_{N_p}}&\ldots&e^{-j\frac{2\pi}{N} P_{N_p}(N-1)}
\end{array}\right].
\end{equation}

According to the periodic structure of the DFT submatrix $\mathbf{F}_P$, the inner product of $\mathbf{f}_l$ and $\mathbf{f}_k$ used in (\ref{eq:12}) only depend on $r=k-l$:
\begin{eqnarray}\label{eq:14}
\tilde{\mu}_{\mathbf{F}_P} = N_p\mu_{\mathbf{F}_P} &=& \max_{1\leq r\leq N-1}   {|\langle \mathbf{F}_{l}\;,\;\mathbf{F}_{l+r}\rangle|} \nonumber\\
&=&\max_{1\leq r\leq N-1}  \Big|\sum_{i=1}^{N_p}e^{-j\frac{2\pi}{N} P_ir}\Big|.
\end{eqnarray}

Here we aim to choose the set $\mathcal{P}$ with $|\mathcal{P}|=N_P$ in order to minimize $\tilde{\mu}$. For the simplicity of analysis, let us define $f(x)=\sum_{i=1}^{N_p}x^{P_i}$. Hence, (\ref{eq:14}) turns into:
\begin{equation}
\label{eq:15}
\tilde{\mu}_{\mathbf{F}_P}=\max_{1\leq r\leq N-1}{\big|f(e^{-j\frac{2\pi}{N} r})\big|}.
\end{equation}

Since we are interested in the modulus values of the function $f(.)$ on the unit circle, instead of $|f(x)|$ it is simpler to work with $|f(x)|^2=f(x)f^* (x)=f(x).f(1/x)$. This shows that the optimum set $\mathcal{P}$ which minimizes the coherence, is found by:
\begin{eqnarray}\label{eq:16}
\mathcal{P}_{opt}&=&
\mathrm{arg}\min_{\mathcal{P}}\max_{1\leq r\leq N-1} f(e^{-j\frac{2\pi}{N} r})f(e^{j\frac{2\pi}{N} r})\nonumber\\
&\equiv& 
\mathrm{arg}\min_{\mathcal{P}}\max_{1\leq r\leq N-1}
\sum_{l=1}^{N_p}\sum_{k=1}^{N_p}e^{-j\frac{2\pi}{N}r (P_l-P_k)}.
\end{eqnarray}

If the set of cyclic differences of $\mathcal{P}$ is defined as $\mathcal{D}= \{ P_k-P_l (\mathrm{mod}\; N) \; \big| \; 1\leq l,k\leq N_p;\,l\neq m\}$ , and $a_d$  denotes the number of repetitions of the number $0\leq d\leq N-1$ in the set $\mathcal{D}$, we have:
\begin{equation}
\sum_{l=1}^{N_p}\sum_{k=1}^{N_p}e^{-j\frac{2\pi}{N}r (P_l-P_k)}
=N_p+\sum_{d=1}^{N-1}a_de^{-j\frac{2\pi}{N}rd}\triangleq g(r,\{a_d\}).
\end{equation}
Therefore, it is clear that we should look for the set of indices $\mathcal{P}$ which minimizes the maximum value of the function $g(r,\{a_d\})$ over all $1\leq r\leq N$. Since
\begin{eqnarray}
\sum_{r=1}^{N-1}g(r,\{a_d\})=N_p(N-1)-\sum_{d=1}^{N-1}a_d = N_p(N-N_p),
\end{eqnarray}
%
it is obvious that
\begin{eqnarray}
\max_{1\leq r\leq N-1} g(r,\{a_d\}) \geq \frac{N_P(N-N_P)}{N-1},
\end{eqnarray}
and the equality happens when
\begin{equation}
g(1,\{a_d\})=\dots=g(N-1,\{a_d\})=\frac{N_P(N-N_P)}{N-1},
\end{equation}
which is valid only for
\begin{equation} a_1=a_2=\dots=a_{N-1}=\frac{\sum_{i=1}^{N-1}{a_d}}{N-1}=\frac{N_p(N_p-1)}{N-1}.
\end{equation}

Hence, if there exists an index set $\mathcal{P}$ for which $a_1=\dots=a_{N-1}$ happens, it is for sure the best possible choice for minimizing the coherence. Such index sets are already known as cyclic difference sets \cite{1IT}; unfortunately, the existence of difference sets are limited to some specific pairs $(N,N_p)$.

\subsection{Greedy Coherence Minimization for Improper Pairs of $N$ and $N_p$} \label{sec:greedys}
Cyclic difference sets described in section \ref{sec:DS} are the optimal choices for the OFDM pilot locations with respect to the coherence criterion. In fact, if we make a DFT submatrix based on a cyclic difference set, the resultant matrix meets the Welch lower bound \cite{welch}. In other words, not only is such a submatrix optimum among all DFT submatrices of its size, but also is the optimum code-book in the sense of minimum coherence among all the matrices with the same size. Nevertheless, for many pairs of $N$ and $N_p$, there is no cyclic difference set. Therefore we have to find proper indices for pilot allocation, using efficient search methods. In \cite{urbanke}, it is suggested to use random exhaustive search among all $N\choose N_p$ DFT submatrices to find the one with the minimum coherence. As $N$ increases, the cardinality of the search space grows exponentially and the results of the random search in relatively small steps might not be satisfactory. Here we propose a greedy method to find suitable pilot index set.

As stated in \ref{sec:DS}, it is important that the set of cyclic differences $\mathcal{D}$ of the set of pilot indices $\mathcal{P}$ has equal number of repetitions ($a_d$) for its different elements; i.e., the variance of the set of repetitions $\{a_d\}_d$ is equal to zero. In our greedy method, we choose $N_p$ pilot indices in the following  $N_p$ stages: since rather than the exact value of the indices, their cyclic difference are important, we initialize the index set by $\mathcal{P}^{(1)}=\{1\}$ ($1$ is arbitrary). The rest of the stages are summarized in the following:\\

For the $i^{th}$ pilot index allocation:
\begin{enumerate}
\item Form all $N-i+1$ possible $i$-element subsets by adding an element to $\mathcal{P}^{(i-1)}$:
\begin{eqnarray}
\{\mathcal{P}^{(i)}\}=\left\{\mathcal{P}^{(i-1)}\cup s \; \big| \; s\in \{1,\dots,N\}\setminus \mathcal{P}^{(i-1)}\right\}
\end{eqnarray}

\item For each $i$-element set generated in step (1), form the set of cyclic differences and the set of repetitions ($\{a_d\}$).

\item Choose the set (or one of the possible sets) with the minimum variance in the elements of the respective repetition set ($\{a_d\}$).

\item If $i<N_p$, go to step (1).
\end{enumerate}

\section{Simulation results}
In order to give an insight toward theoretical results presented in sections \ref{sec:IMAT} and \ref{sec:Pilot_coherence}, we conduct several computer simulations using MATLAB. The simulations are presented in two parts:

\subsection{IMAT Method for Sparse Channel Estimation Results}
The iterative method with adaptive thresholding presented in the block diagram of Fig. 1 is simulated in an OFDM system based on DVB-H standard with slight modifications. All the parameter specifications are presented in Table \ref{tab:1}. For the channel, we have considered a Rayleight multipath fading channel with $4$ significant nonzero taps at normalized (to carrier spacing) doppler frequency of $1\%$; the average delay and power of the taps are presented in Table \ref{tab:Aaa}.
\begin{table}[th]
\caption{simulation parameters}
\centering
\begin {tabular}{c c}
\hline\hline
Parameter&Specifications\\[0.5ex]
\hline
Number of Subcarriers 	& 	$256$\\
Number of Pilots 			& 	$16(6.25\%)$\\
OFDM Symbol Duration 	&  $224(\mu s)$\\
Cyclic Prefix Length		& 	$32(1/8)$\\[1ex]
\hline
\end{tabular}
\label{tab:1}
\end{table}\\
\begin{table}[th]
\caption{Fading Channel Parameters}
\centering
\begin {tabular}{c c c c c}
\hline\hline
Delay($\mu s$)& $1.7$ & $3.5$ & $5.2$ & $11.3$ \\
\hline
Power(dB)& $-2$ & $0$ & $-5$ & $-7$ \\
\hline
\end{tabular}
\label{tab:Aaa}
\end{table}

\begin{figure}[!t]
\centering
\includegraphics[totalheight=5.5cm,width=9cm]{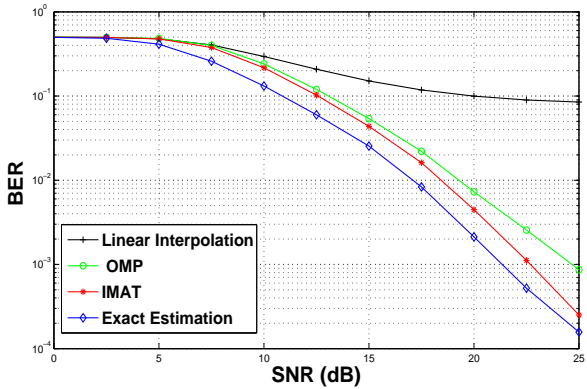}
\caption{BER of different estimators at various SNRs}
\label{fig:ber}
\end{figure}

\begin{figure}[!t]
\centering
\includegraphics[totalheight=5.5cm,width=9cm]{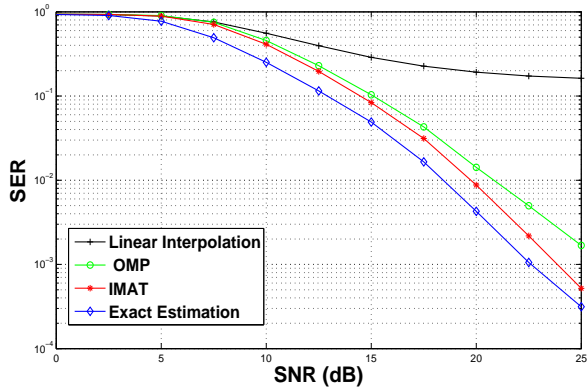}
\caption{SER of different estimators at various SNRs}
\label{fig:ser}
\end{figure}

The IMAT method is compared with the linear interpolation method which estimates the channel at pilot frequencies using LS estimate (2) and then uses a linear interpolation function to estimate the CFR at data subcarriers. Also orthogonal matching pursuit is simulated as a proper sparse reconstruction method for channel estimation. The obtained curves of the obtained Bit Error Rate (BER) and Symbol Error Rate (SER) 
shown in Fig. \ref{fig:ber} and \ref{fig:ser} indicate that the IMAT method outperforms the other competitors.

\subsection{Pilot Allocation in Sparsity-Based Estimation Methods}
\begin{figure}[!t]
\centering
\includegraphics[totalheight=5.5cm,width=9cm]{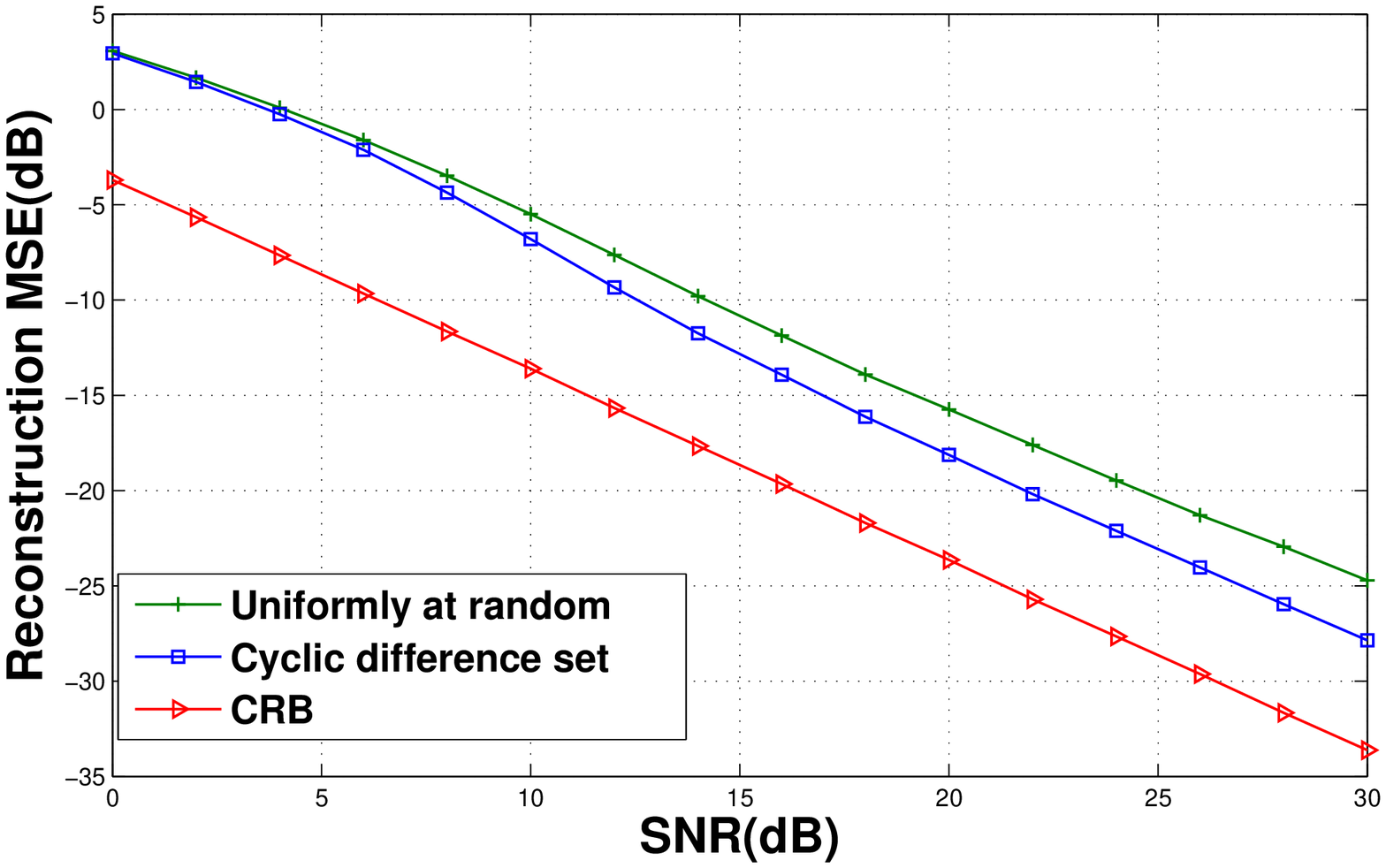}
\caption{MSE of proposed and random pilot allocation methods for OMP reconstructuon.}
\label{fig:mseomp}
\end{figure}

\begin{figure}[!t]
\centering
\includegraphics[totalheight=5.5cm,width=9cm]{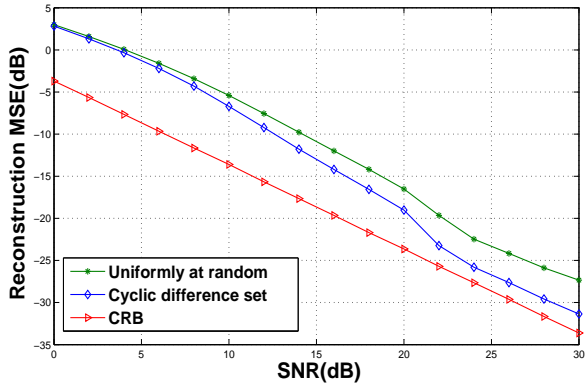}
\caption{MSE of the proposed and random pilot allocation methods for IMAT reconstruction.}
\label{fig:mseimat}
\end{figure}

In this part, we compare the MSE and perfect reconstruction percentage in channel estimation for pilot allocation methods presented in this paper. For our simulations in this part, we generated a random 3-tap channel with varying fading parameters in each OFDM block and averaged the results over $5000$ runs. Fig. \ref{fig:mseomp} shows the MSE of the estimated channel for two different methods of pilot allocation. In the first scenario, the pilots are chosen uniformly at random for each block; in our proposed scheme, the pilots are arranged according to a $(73,9,1)$ cyclic difference set and its cyclic shifts for different OFDM blocks. The Cramer-Rao lower bound is also presented in the figures to give us a meaningful goal standard. This bound is given by \cite{2-Nd}:
\begin{equation}
CRB=\sigma^2\;\mathrm{trace}\left(\left({{\mathbf{F}_{p,\Lambda}}^H\mathbf{F}_{p,\Lambda}}\right)^{-1}\right),
\end{equation}
where $\sigma^2$ is the noise variance and $\mathbf{F}_{p,\Lambda}$ is the submatrix of $\mathbf{F}_p$ obtained by keeping the columns corresponding to the channel taps. Cramer-Rao bound is a fair criterion to measure the quality of our pilot allocation method, since it is the MSE of an estimator that knows the exact location of the channel taps. Therefore, in OMP channel estimation, if we select pilot locations properly, the maximum cross correlation between the columns of $\mathbf{F}_p$ becomes small. Hence, the columns of the resultant measurement matrix in (3) become less correlated which makes it easier to detect and estimate the CIR. Similarly for IMAT, decrease in the off-diagonal elements of the distorting matrix results in the decrease of the MSE values of the estimated channel (Fig. \ref{fig:mseimat}).

\begin{figure}[!t]
\centering
\includegraphics[totalheight=5.5cm,width=9cm]{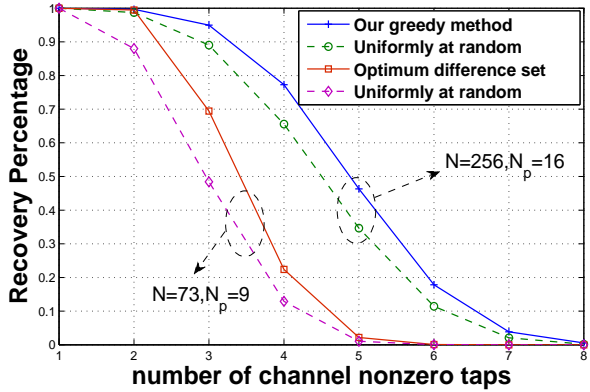}
\caption{Recovery percentage for the proposed \& random pilot allocation schemes}
\label{fig:percent}
\end{figure}

Finally, we compare successful channel recovery percentage in noiseless case for different pilot allocation methods. For this purpose, we have considered the OFDM communication with $N=256$ subcarriers and $N_p=16$ pilots. Since there is no cyclic difference set for this pair of $N$ and $N_p$, we have employed the proposed greedy search in section \ref{sec:greedys} to find a pattern for pilots with small coherence. The recovery percentage which is the percentage of exact channel recovery without error for various number of channel taps and OMP reconstruction method is presented in Fig. \ref{fig:percent}.

\section{Conclusion}\label{sec:conc}

In this paper, we investigated the problem of OFDM pilot allocation in sparsity-based channel estimation methods. First, we proposed an iterative method with adaptive thresholding (IMAT) which detects channel nonzero taps and their corresponding values iteratively for the purpose of OFDM channel estimation. As it was shown in the simulation results, this method outperforms typical interpolation methods such as Linear Interpolation and greedy algorithms in sparse channel estimation. We derived the optimum pilot location for greedy methods in sparse channel estimation, based on minimizing the coherence in DFT submatrices. Simulation results show the improvement in the MSE of the estimated channel for our proposed pilot allocation method compared to uniformly random insertion of pilots.



\end{document}